\documentstyle[12pt,aps]{revtex}
\begin{document}
\draft
\title{\large {\bf The proton and neutron distributions in Na
isotopes: \\
the development of halo and shell structure}}
\author{J.Meng, I.Tanihata and S.Yamaji}
\address{ The Institute of Physical and Chemical Research (RIKEN), 
Hirosawa 2-1, Wako-shi\\
Saitama 351-01,  JAPAN}
\date{\today}
\maketitle
\begin{abstract}
The interaction cross sections for $^A$Na $+$
$^{12}$C reaction are calculated using Glauber model. 
The continuum Hartree-Bogoliubov theory has been generalized 
to treat the odd particle system and take the continuum 
into account.
The theory reproduces 
the experimental result quite well. 
The matter distributions from the proton drip line to the 
neutron drip line in Na isotopes have been systematically 
studied and presented. 
The relation between the shell effects and the halos has
been examined. The tail of the matter distribution 
shows a strong dependence on the shell structure. 
The neutron $N=28$ closure shell fails to appear due to the 
coming down of the $2p_{3/2}$ and $2p_{1/2}$.
The development of the halo was understood as changes in 
the occupation in the next 
shell or the sub-shell close to the continuum limit. 
The central proton density is found to be decreasing near the 
neutron drip line, which is due to the proton-neutron interaction.
However the diffuseness of the proton density does not 
change for the whole Na isotopes.
\end{abstract}

      
\pacs{Key word: Relativistic mean field, pairing, continuum, 
                halo,shell structure \\}
\par
The recent developments in the accelerator technology and the
detection techniques all around the world 
have changed the nuclear
physics scenario. It is now possible to produce and study the
nuclei far away from the stability line -- so called "EXOTIC
NUCLEI". Experiments of this kind have casted new light on
nuclear structure and novel and entirely unexpected
features appeared: e.g. the neutron halo in $^{11}$Li \cite{THH.85b} 
and neutron skin\cite{Alk.97} as the rapid increase 
in the measured interaction cross-sections in the
neutron-rich light nuclei. The extreme proton and neutron 
ratio of these nuclei and physics
connected with these low density matter have 
attracted more and more attentions 
in nuclear physics as well as other fields such as astrophysics.

With the exotic matter distribution 
near the drip line, a lot of questions are still open, e.g., 
the relation between the halo and the shell effect , 
the difference about the proton and neutron 
distribution on the stability line and away from the stability line.
How is the halo formed ? Are there a rapid change from the normal 
nuclear density to the halo density or 
a gradual change in the particle number ?  As the matter 
distribution is not measurable directly , series of experiment at different 
incident beam energy are necessary in order to determine the density 
distribution of both proton and neutron model-independently.  
Among all the experiments
carried out so far, Na isotopes provide a good opportunity
to study the density distributions over a wide range of neutron
numbers \cite{Suz.95}. Although theoretically, 
lots of works have been reported with either the non-relativistic 
Hartree-Fock or 
relativistic mean field, but the pairing property is always 
neglected or simply treated by the BCS approximation, e.g., see 
\cite{Cam.75},\cite{Pat.91}.  
For the pairing correlation, the contribution from the continuum 
are essential for the description of the 
drip line nuclei.
We report in this
letter a systematic study of nuclear density distributions in Na isotopes
within the Relativistic Mean Field Theory (RMF), with the pairing 
and the blocking effect for odd particle system 
properly described by Hartree-Bogoliubov theory in 
coordinate representation, 
and try to answer
some of the general questions in these low-density nuclear matter.

Definitely 
it is necessary to have a self-consistent theory to describe 
directly the cross section, so that the usual model dependent 
way for extracting the matter distribution could be avoided.
As discussed in great detail in a recent review article
\cite{ZHU.93} and in the references given therein, one has
applied so far rather different techniques to describe
halo phenomena in light nuclei, as for instance the exact
solution of few-body equations treating inert sub-clusters
as point particles,  or the density dependent Hartree-Fock
method in a localized mean field taking into account all
the particles in a microscopic way, or a full shell model
diagonalizations based on oscillator spaces with two
different oscillator parameters for the core- and the
halo-particles.
Recently, a fully self-consistent calculation within
the Relativistic Hartree-Bogoliubov (RHB) theory in coordinate space
for the description of the chain of Lithium isotopes ranging
from $^{6}$Li to $^{11}$Li was reported \cite{MR.96}.
It combines the advantages of a proper description of
the spin-orbit term with those of full Hartree-Bogoliubov
theory in the continuum, which allows in the canonical
basis the scattering of Cooper pairs to low lying resonances
in the continuum. 
Excellent agreement including binding 
energy, matter radius, and matter distribution with the
experimental data was obtained without any modification of the
neutron $1p_{1/2}$ level like the former works
( see  \cite{MR.96} and the references therein ).

We first study the isospin
dependence of the density distribution and the ground
state properties of the Na nuclei within RHB. 
Then the cross sections based on Glauber model calculations 
with the density obtained from RHB 
are directly compared with the experimentally determined
ones.
As the theory used here is fully microscopic 
and parameter free, it gives consistent description 
of the proton and neutron distribution, and the development of 
proton and neutron halo or skin could be examined.

The basic ansatz of the RMF theory is a Lagrangian
density whereby nucleons are described
as Dirac particles which interact via the exchange of various 
mesons and the photon.
The mesons considered are the scalar
sigma ($\sigma$), vector omega ($\omega$) and iso-vector vector rho
($\rho$).  The latter provides the necessary isospin
asymmetry\cite{RI.96}.
The scalar sigma meson moves in a self-interacting field having cubic and
quartic terms with strengths $g_2$ and $g_3$ respectively.
The Lagrangian then consists of the free baryon
and meson parts and the interaction part with minimal coupling,
together with the nucleon mass $M$ and $m_\sigma$
($g_\sigma$), $m_\omega$ ($g_\omega$) and $m_\rho$ ($g_\rho$) the masses
(coupling constants) of the respective mesons:
\begin{equation}
\begin{array}{rl}
{\cal L} &= \bar \psi (i\rlap{/}\partial -M) \psi +
\,{1\over2}\partial_\mu\sigma\partial^\mu\sigma-U(\sigma)
-{1\over4}\Omega_{\mu\nu}\Omega^{\mu\nu}\\
\ &+ {1\over2}m_\omega^2\omega_\mu\omega^\mu
-{1\over4}{\vec R}_{\mu\nu}{\vec R}^{\mu\nu} +
 {1\over2}m_{\rho}^{2} \vec\rho_\mu\vec\rho^\mu
-{1\over4}F_{\mu\nu}F^{\mu\nu} \\
 &-  g_{\sigma}\bar\psi \sigma \psi~
     -~g_{\omega}\bar\psi \rlap{/}\omega \psi~
     -~g_{\rho}  \bar\psi 
      \rlap{/}\vec\rho
      \vec\tau \psi
     -~e \bar\psi \rlap{/}A \psi
\end{array}
\end{equation}
where a non-linear scalar self-interaction $U(\sigma)$
of the $\sigma$ meson 
has been taken into account\cite{BB.77}. 

For the proper treatment of the 
pairing correlations and for correct description of 
the scattering of Cooper pairs
into the continuum  in a 
self-consistent way, one needs to extend the present relativistic
mean-field theory to a continuum
RHB theory\cite{DFT.84}.
Using Green's function techniques it has been shown in Ref.
\cite{KR.91} how to derive the RHB
equations from such a Lagrangian:
\begin{equation}
\left(\begin{array}{cc} h & \Delta \\ -\Delta^* & -h^* \end{array}\right)
\left(\begin{array}{r} U \\ V\end{array}\right)_k~=~
E_k\,\left(\begin{array}{r} U \\ V\end{array}\right)_k,
\label{RHB}
\end{equation}
$E_k$ are quasi-particle energies and the coefficients
$U_k(r)$ and $V_k(r)$ are four-dimensional Dirac spinors.
$h$ is the usual Dirac Hamiltonian
\begin{equation}
h~=~\mbox{\boldmath $\alpha p$}~+~g_\omega\omega~+~ \beta(M+g_\sigma
\sigma)~-~\lambda
\label{h-field}
\end{equation}containing the chemical potential $\lambda$ adjusted to the
proper particle number and the meson fields $\sigma$ and
$\omega$ determined as  usual in a self-consistent way from
the Klein Gordon equations in {\it no-sea}-approximation.

The pairing potential $\Delta$ in Eq. (\ref{RHB}) is given
by
\begin{equation}
\Delta_{ab}~=~\frac{1}{2}\sum_{cd} V^{pp}_{abcd} \kappa_{cd}
\label{gap}
\end{equation}
It is obtained from the pairing tensor $\kappa=U^*V^T$ and
the one-meson exchange interaction $V^{pp}_{abcd}$ in the
$pp$-channel. 
As in Ref.
\cite{MR.96} $V^{pp}_{abcd}$ in Eq.
(\ref{gap}) is the density dependent two-body force of zero
range:
\begin{equation}
V(\mbox{\boldmath $r$}_1,\mbox{\boldmath $r$}_2)
~=~\frac{V_0 }{2}(1+P^\sigma)
\delta(\mbox{\boldmath $r$}_1-\mbox{\boldmath$r$}_2)
\left(1 - \rho(r)/\rho_0\right).
\label{vpp}
\end{equation}

The RHB equations (\ref{RHB}) for zero range pairing forces
are a set of four coupled differential equations for the HB
quasi-particle Dirac spinors $U(r)$ and $V(r)$. They are solved 
by the shooting
method in a self-consistent way as \cite{MR.96}. 
With a step size of $0.1$ fm and using
proper boundary conditions the above equations are solved
in a spherical box of radius $R = 25 $ fm.  As shown in 
\cite{MR.97}, for these relatively lighter nuclei $R = 25 $ fm
give quite accurate result.
Since we use a pairing force of zero range
(\ref{vpp}) we have to limit the number of continuum levels
by a cut-off energy.  For each spin-parity channel 20
radial wave functions are taken into account, which
corresponds roughly to a cut-off energy of
120 MeV for $R=25$ . After fixing the cut-off energy and the box radius
$R$, the strength $V_0$ of the pairing force (\ref{vpp}) for
both the neutrons and protons is 
fixed by a similar calculation of Gogny force D1S \cite{BGG.84} by 
reproducing the corresponding pairing energy
$-\frac{1}{2}\mbox{Tr}\Delta\kappa$i as ref.\cite{MR.96}. 
For $\rho_0$ we use the nuclear
matter density 0.152 fm$^{-3}$. 
The ground state $|\Psi>$ of the even particle system
is defined as the vacuum with respect to the quasi-particle:
$\beta_{\nu} |\Psi >=0$, $|\Psi > = \prod_\nu \beta_{\nu} |->$, 
where 
$|->$ is the bare vacuum. For odd system, the ground state
can be correspondingly written as:
$|\Psi >_{\mu} = \beta_{\mu}^\dagger\prod_{\nu \ne \mu} \beta_{\nu} | - >$, 
where $\mu$ is the level which is blocked. 
The exchange of the quasiparticle creation operator
$\beta_{\mu}^\dagger$ with the corresponding annihilation
operator $\beta_{\mu}$ means the replacement of the column
$\mu$ in the  $U$ and $V$ matrices by the corresponding
column in the matrices $V^*$, $U^*$ \cite{RS.80}.

A systematic set of calculations have been carried out for all the
nuclei in Na isotopes with mass number A ranging
from 17 to 45.
We have employed in the calculations the non-linear Lagrangian 
parameter set NLSH
which was widely used for the description of all the 
medium and heavy nuclei, particularly drip line nuclei\cite{SNR.93}. 

The calculated binding energies $E_B$ and the interaction 
cross sections with the Glauber Model are presented in Figure 1.
The calculated binding energies $E_B$ are in good agreement with
the empirical values\cite{AUD.93} for the radioactive isotopes, 
which are our current interests.
The resonance states of 
$^{17}$Na and  $^{18}$Na ( with a positive Fermi energy ) 
are exactly reproduced.
$^{19}$Na is bound but unstable against the proton emission, 
reproducing the experimental observation. The neutron 
drip-line nucleus has been predicted to be $^{45}$Na in the present 
model. The difference between the calculations and the empirical values
for the stable isotopes is from the deformation, which has been neglected 
here.

To compare the cross section directly with experimental 
measured values, 
the densities $\rho_{n,p}(r)$ of the target $^{12}$C and 
the Na isotopes obtained from RHB ( see Fig.2 ) 
were used. The cross sections were calculated 
in Glauber model by using 
the free nucleon-nucleon cross section\cite{Ray.79} 
for the proton and neutron respectively. 
The cross sections for reaction of Na isotopes 
at $950A$ MeV on $^{12}$C have been compared with the experimental 
values\cite{Suz.95} in the upper part of Fig.1. 
The agreement between the calculated results and measured
ones are fine. The cross section below $^{22}$Na changes only 
slightly with the neutron number, which 
means the proton density has played important role to remedy the 
contribution of less neutron. From $^{25}$Na 
to the neutron drip line, a gradual increase of
the cross section has been observed. After $^{32}$Na, although 
no data exist yet, a 
relatively fast increase has been predicted. As we have seen 
here, the measured cross section shows similar behavior with 
that of RHB. 

As the density, which is obtained from a fully microscopic and 
parameter free model,  is well supported by the experimental 
cross sections and binding energies,
we proceed to examine the density distributions 
of the whole isotopes and study the relation between 
the development of halo and shell effect within the model.

The density distributions for both the proton and the neutron 
are given in Fig.2 .  As seen in the upper part of Fig.2 , the change of the
neutron density is as follows: for the nucleus with less number of neutrons 
(N), the density at the center is low and it spreads 
only to some smaller distance. With the increase of N,
the density near the center increases due to the occupation of 
the $2s_{1/2}$ level, so does the 
development of neutron radius. In the lower part of Fig.2 , 
the proton density shows different behavior. The surface 
is more or less unchanged because of the Coulomb Barrier, 
but with the increase of N,
the density of the center 
decreases due to the slight increasing at the tail. 
This is considered to be due to the attractive 
proton neutron interaction. But as 
the density must be multiplied by a factor 
$4\pi r^2$ before the integration in order 
to give the fixed Z, the big change in 
the center does not influence the outer part of the proton 
distribution very much. This result is 
fully consistent with the recent experiment on 
the charge-changing cross section\cite{Bo.97}.

The matter densities for the even N 
Na isotopes are given in Fig.3. 
the shortest tail
in the  total density
occurs for $^{23-25}$Na, the most stable ones. For either the
proton or the neutron rich side, the tail density increases
monotonically. The tail ( $r > 10$ fm ) of the proton rich nuclei 
is mainly due to the contribution of the proton and
that of the neutron drip-line is mainly due to the contribution of the
neutron. Compared with the neutron-rich isotopes, 
the proton distribution with less N 
has higher density at the 
center, lower density in the middle ( $2.5 < r < 4.5 $fm ),
a larger tail in the outer ( $r > 4.5 $ fm ) part.

Next we will examine the density distribution for the neutron 
rich side. It is interesting to connect the matter distribution 
with the level distribution ( See Fig.4 ): 
after $^{25}$Na, as it is a sub-closure shell for 
the $1d_{5/2}$, then the
neutrons are filled in the $2s_{1/2}$ and $1d_{3/2}$. 
So the tail of the density for $^{27}$Na is two 
order of magnitude larger than $^{25}$Na at $r = 10$ fm, 
while the tail of the density from 
$^{27}$Na to $^{31}$Na is very close to each other. 
But as more neutrons are filled in, the added neutrons are filled in
the next shell $1f_{7/2}$, $2p_{3/2}$ and $2p_{1/2}$. 
So again two order of magnitude's increase has been seen 
from $^{31}$Na to $^{33}$Na, and then a gradual increase 
after $^{33}$Na. So it becomes clear that the rapid increase 
in the cross section is connected with the filling of 
neutrons in the next shell or sub shell. In 
the inserts in Fig.3 : the radius $r_0$ at which 
$\rho(r_0) = 10^{-4}$ fm$^{-3}$ is given as a function of the mass number 
to see the relation between the shell effect and matter 
distribution more clearly.
It is very interesting to see a slight decrease of 
$r_0$ from proton drip-line to $^{25}$Na. 
The tail of $^{25}$Na is the smallest. From $^{26}$Na to 
$^{31}Na$, one sees a almost constant $r_0$.
After a jump from $^{31}$Na to $^{32}$Na,
a rapid increasing tendency appears again.

In Figures 4 , the 
microscopic structure of the 
single particle energies in the canonical
basis \cite{RS.80} is given.
In the left panel of Fig. 4, 
the single particle levels in the canonical basis for the
isotopes with an even neutron number are shown.
Going from $A=19$ to $A=45$ we observe a big gap
above the $ N=8 $, $ N=20 $ major shell , and $ N=14 $ sub-shell. 
The $N=28$ shell for stable nuclei fails to appear, as the $2p_{3/2}$
and $2p_{1/2}$ come so close to $1f_{7/2}$.
When $N \ge 20 $, the neutrons are filled to the
levels in the continuum or weakly bound states in the order
of $1f_{7/2}$, $2p_{3/2}$, $2p_{1/2}$, and  $1f_{5/2}$. 
In the right part, 
the occupation probabilities in the
canonical basis of all the neutron levels 
below  $E = 10$ MeV  have been given 
for $^{35}$Na to show how the levels are filled in 
nuclei near the drip line. The importances of 
careful treatment of the pairing correlation, of 
treating properly the scattering of particle pairs to 
higher lying levels, are noted in the figure.

Summarizing our investigations, the 
development of a proton skin as well as neutron skin 
has been systematically studied 
with a microscopic RHB model, where the pairing and blocking 
effect have been treated self-consistently.
A systematic set of calculations for the ground state properties
of nuclei in Na isotopes is presented using the
RHB together with standard Glauber theory. 
The RHB equations are solved
self-consistently in coordinate space so that the 
continuum and the pairing have been better treated. 
The calculated binding energies
are in good agreement with the experimental values.
A Glauber model calculation
has been carried out with the density obtained from RHB.
A good agreement has been obtained with the measured
cross sections 
for $^{12}$C as a target and a rapid 
increase of the cross sections has been predicted for 
neutron rich Na isotopes beyond $^{32}$Na.
The systematics of the proton and neutron distribution
are presented. 
After systematic examination of 
the neutron, proton and matter distributions
in the Na nuclei from the proton drip-line
to the neutron drip-line, 
the connection between the tail part of the density and the 
shell structure has been found.
The tail of the matter distribution is not so sensitive 
to how many particles are filled in a major shell. Instead 
it is very sensitive to whether this shell has occupation or 
not.
The physics behind the skin and halo has been revealed as 
a spatial demonstration of shell effect: simply the extra 
neutrons are filled in the next shell and sub-shell. This is in 
agreement with the mechanism observed so far in the halo system but 
more general. As the $1f_{7/2}$ is very close to the continuum, 
the $N=28$ close shell for stable nuclei fails to appear due to the
coming down of the $2p_{3/2}$ and $2p_{1/2}$ levels. 
Another important conclusion here is that, 
contrary to the usual impression, the proton density distribution 
is less sensitive to the proton and neutron ratio. Instead 
it is almost unchanged from the proton drip-line to the neutron  
drip-line. Similar conclusion has been obtained recently 
by charge change reaction experimently\cite{Bo.97}. 
The influence of the deformation, 
which is neglected in the present investigation, 
is also interesting to us,
more extensive study by extending the present study to 
deformed cases are in progress



\leftline{\Large {\bf Figure Captions}}
\parindent = 2 true cm
\begin{description}

\item[Fig. 1]
Upper part: The interaction cross sections
$\sigma_I$  of $^A$Na isotopes on a
carbon target at $950 A$ MeV: the open circles are the result of 
RHB and the available experimental data ( $A = 20 - 23, 25 - 32$ ) 
are given by solid 
circles with their error-bar. The dashed line is a simple 
extrapolation based on the RHB calculation for $^{28-31}$Na.
Lower part: 
Binding energies for Na isotopes, the convention is the same as the upper
part, but the RHB result for particle unstable isotopes are 
indicated by triangle.

\item[Fig. 2]
The neutron ( upper ) and proton ( lower ) density distributions
in Na isotopes. The same figures but in logarithm scale are given 
as inserts to show the tail part of the density distribution.

\item[Fig. 3]
The same as Fig.2 but for matter density distribution. The upper part 
is given in logarithm scale ( the radius is labeled 
at the top of the figure ) and the radius $r_0$ at which 
$\rho(r_0) = 10^{-4}$ for different isotopes is given as inserts.

\item[Fig. 4]
Left part: Single
particle energies for neutrons in the canonical basis as a
function of the mass number. The dashed line indicates the
chemical potential.
Right part: The occupation probabilities in the canonical
basis for $^{35}$Na.

\end{description}
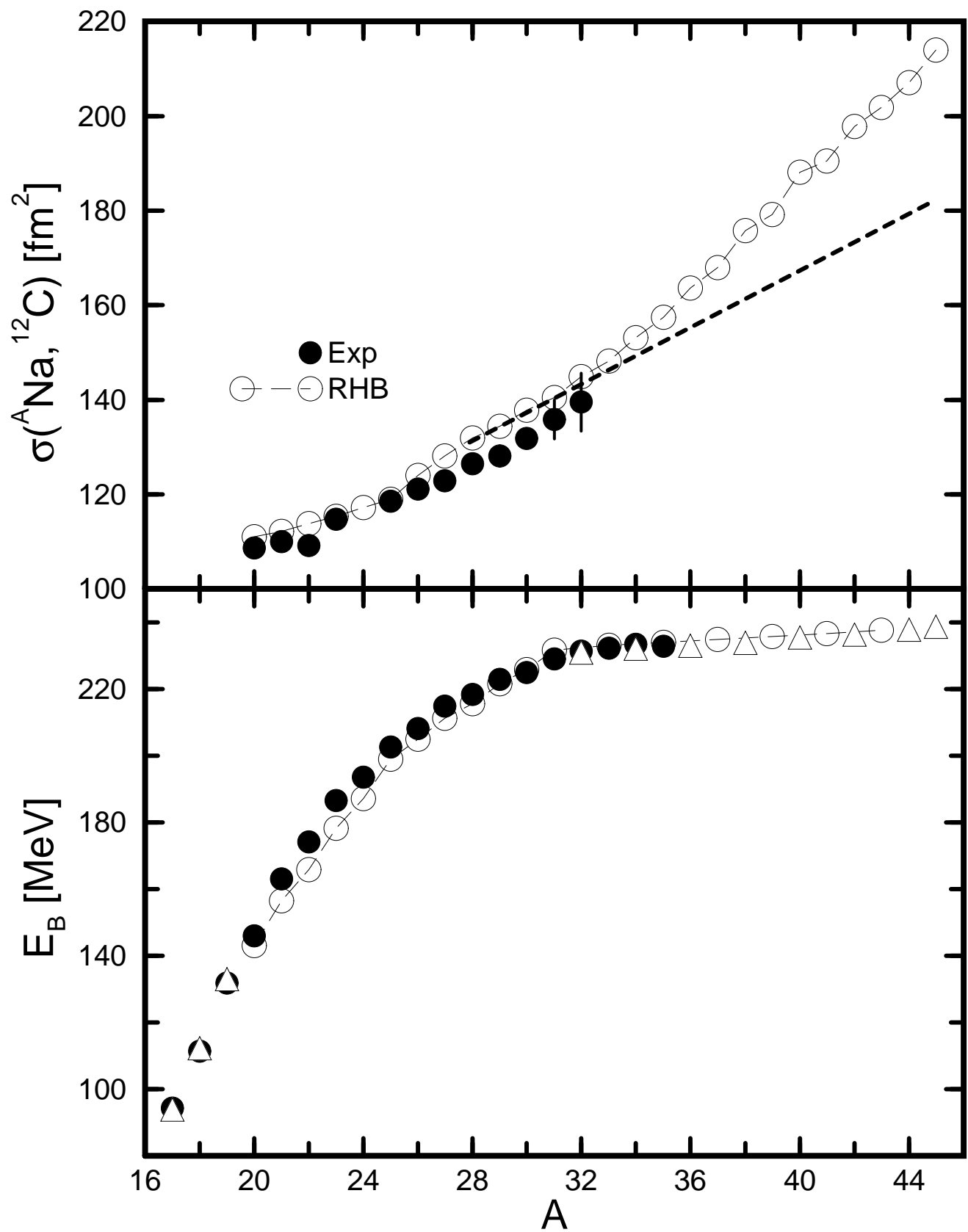
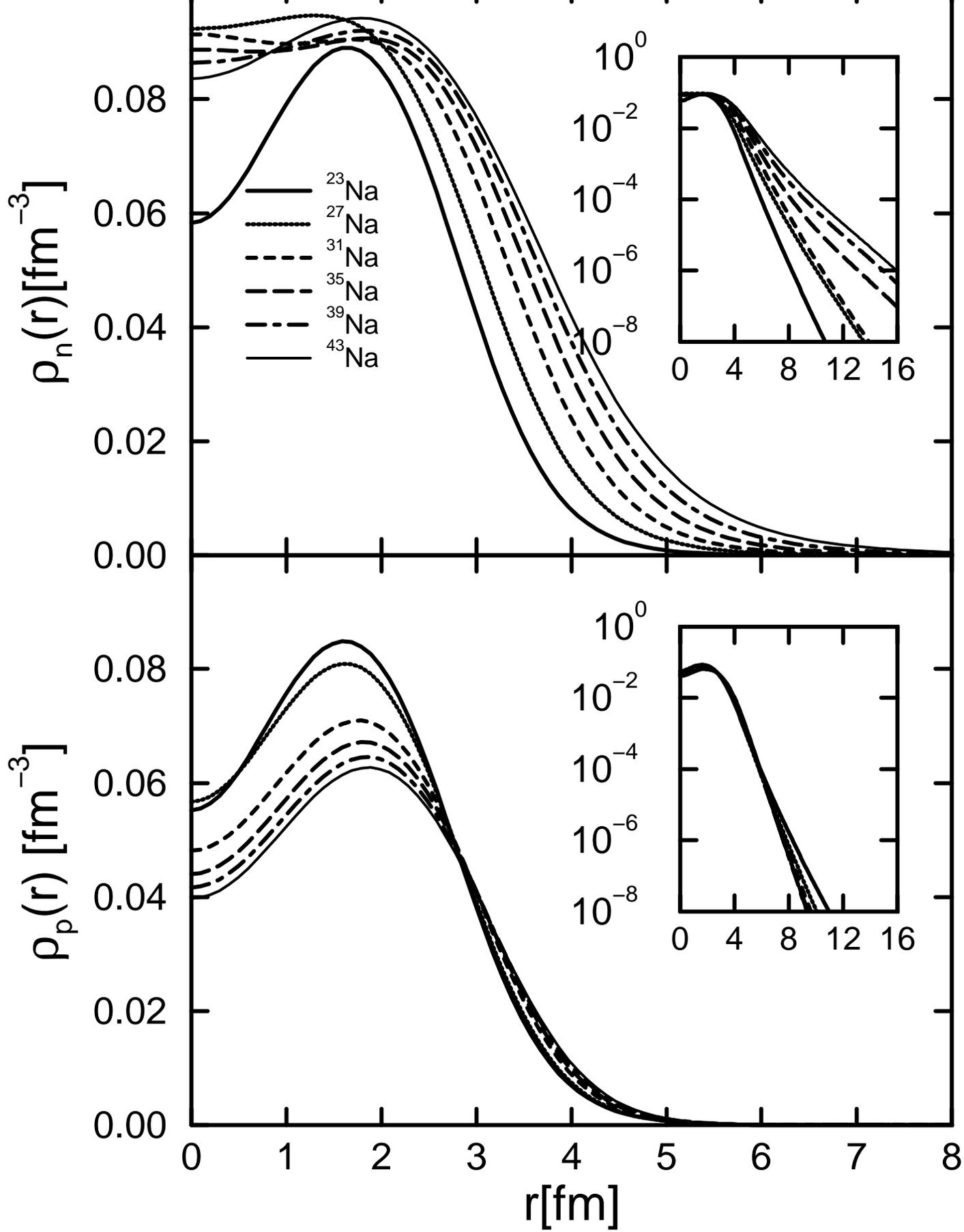
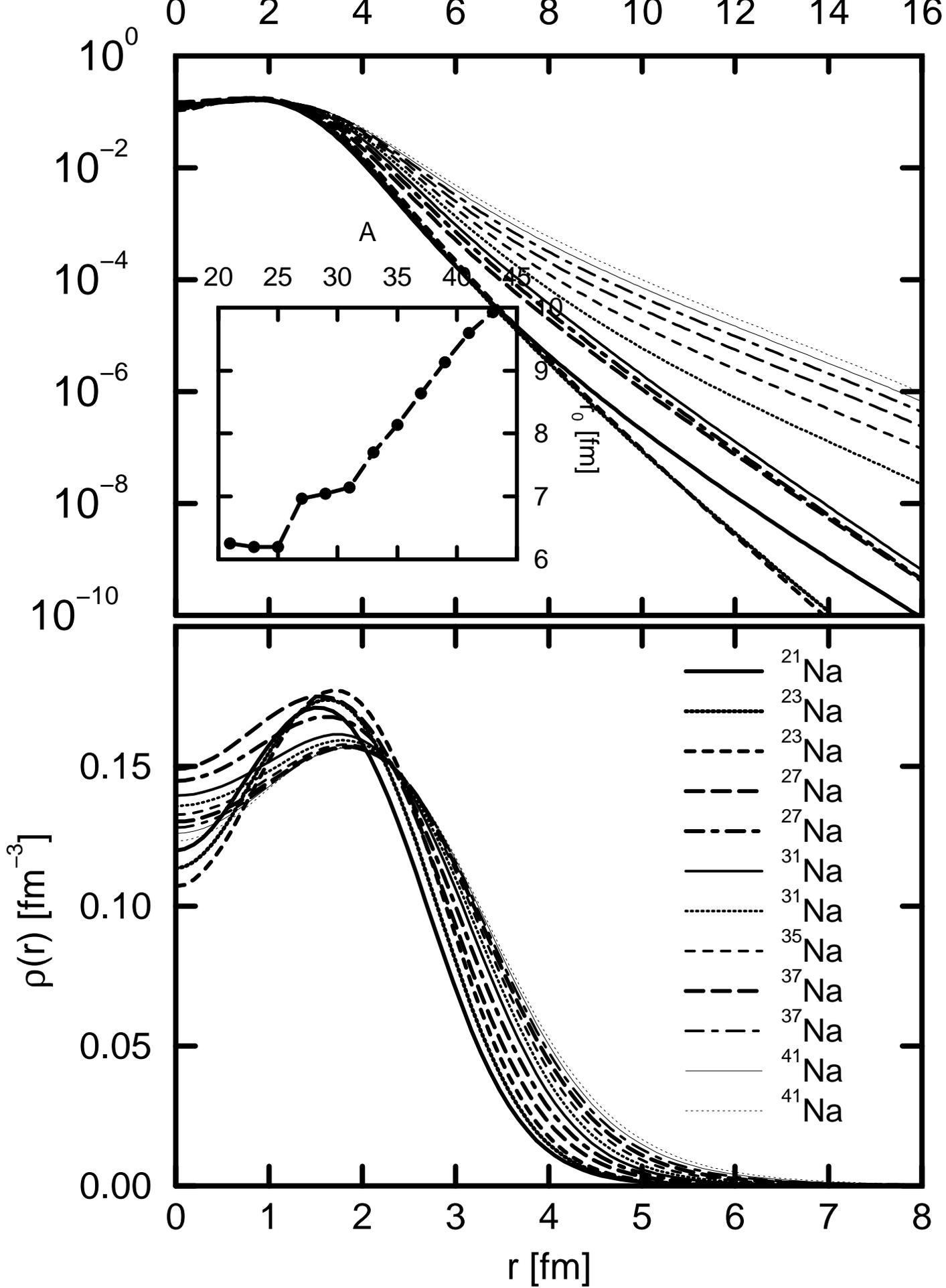
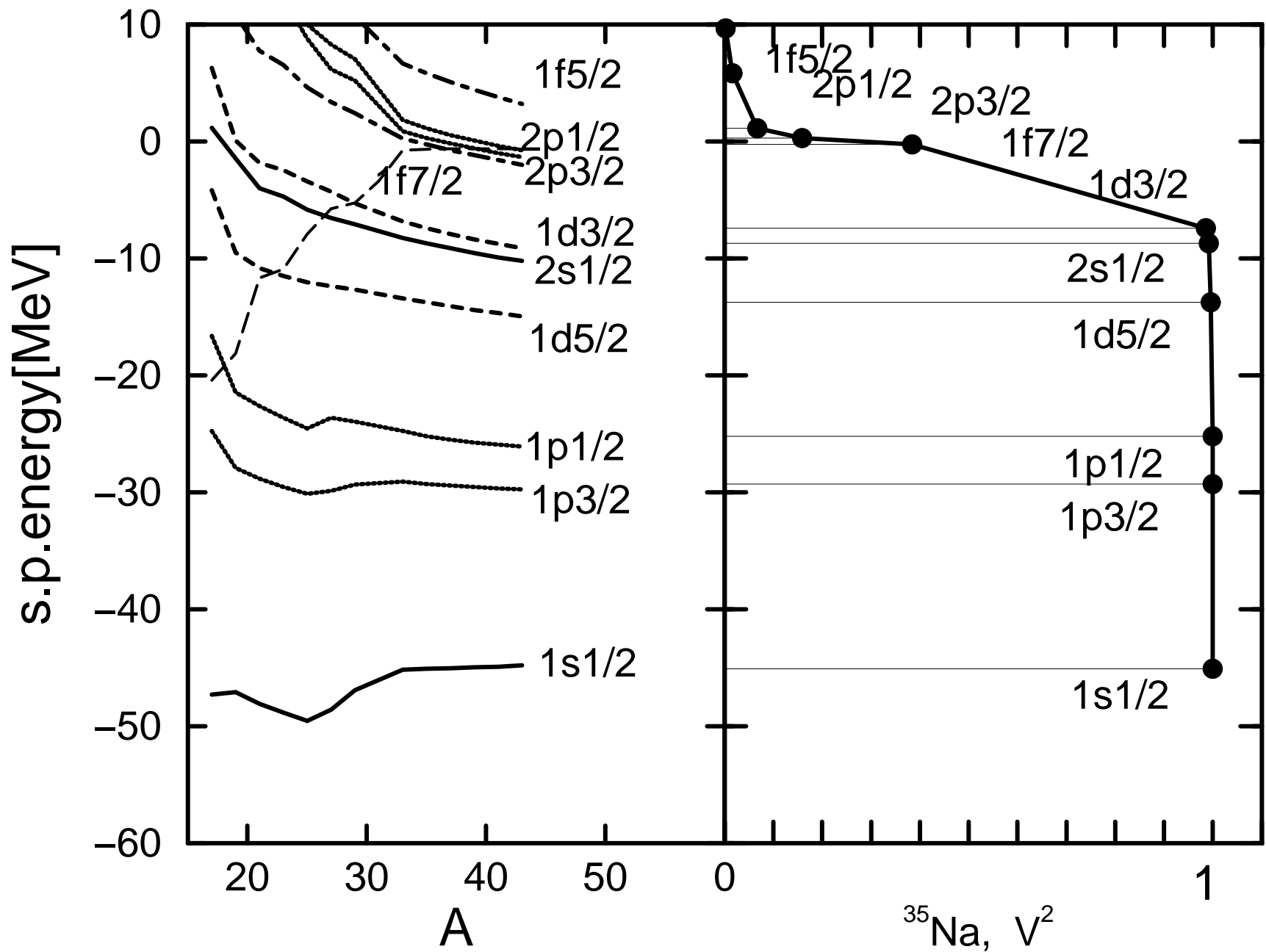
\end{document}